\begingroup

\PassOptionsToPackage{pdfpagelabels=false}{hyperref}
\documentclass[fleqn,usenatbib]{mnras}
\usepackage{newtxtext,newtxmath}

\usepackage[T1]{fontenc}
\usepackage{ae,aecompl}
\usepackage[mathletters]{ucs}
\usepackage[utf8x]{inputenc}
\usepackage{cleveref}

\usepackage{graphicx}	
\usepackage{amsmath}	
\usepackage{amssymb}	

\newcommand{\CO}{$^{12}$CO\ }
\newcommand{\kms}{km/s}

\defcitealias{FFtech}{Paper~I}
\defcitealias{FFredshift}{Paper~II}
\defcitealias{FFlineID}{Paper~III}
\defcitealias{FFncc}{Paper~IV}

\title[SPIRE FF II: Radial Velocity]{The \textit{Herschel} SPIRE Fourier Transform Spectrometer Spectral Feature Finder II. 
Estimating Radial Velocity of SPIRE Spectral Observation Sources\thanks{\textit{Herschel} was an ESA space observatory with science instruments provided by European-led Principal Investigator consortia and with important participation from NASA.}}

\author[J. P. Scott et al.]{Jeremy P. Scott$^{1}$\thanks{E-mail: jeremy.scott@uleth.ca},
Natalia H\l{}adczuk$^{2,3}$,
Locke D. Spencer$^{1}$,
Ivan Valtchanov$^{4}$,
\newauthor
Chris Benson$^{1}$,
Rosalind Hopwood$^{4,5}$
\\
$^{1}$Department of Physics \& Astronomy, University of Lethbridge, AB, T1K 3M4, Canada\\
$^{2}$ European Space Astronomy Centre, ESA, Camino Bajo del Castillo, 28692, Villanueva de la Ca$\tilde{n}$ada, Madrid, Spain \\
$^{3}$ Gran TeCan, S.A., Instituto de Astrof\'{i}sica de Canarias, C/ V\'{i}a L\'{a}ctea, S/N, 38205 - San Crist\'{o}bal de La Laguna, S/C de Tenerife, Spain\\
$^{4}$Telespazio Vega UK for ESA, European Space Astronomy Centre, Operations Department, 28691 Villanueva de la Ca\~nada, Spain\\
$^{5}$Department of Physics, Imperial College London, Prince Consort Road, London SW7 2AZ, UK \\
}

\date{Accepted 2020 June 03. Received 2020 June 03; in original form 2020 February 27}

\pubyear{2020}

\begin{document}
\label{firstpage}
\pagerange{\pageref{firstpage}--\pageref{lastpage}}
\maketitle

\begin{abstract}
The \emph{Herschel} SPIRE FTS Spectral Feature Finder (FF) detects significant spectral features within SPIRE spectra and employs two routines, and external references, to estimate source radial velocity. The first routine is based on the identification of rotational \CO emission, the second cross-correlates detected features with a line template containing most of the characteristic lines in typical far infra-red observations. In this paper, we outline and validate these routines, summarise the results as they pertain to the FF, and comment on how external references were incorporated.  
\end{abstract}




\section{Introduction}

The Herschel Space Observatory (\textit{Herschel}; \cite{Pilbratt10}) was launched on May 14, 2009 as a cornerstone mission of the European Space Agency's (ESA) Horizon 2000 long-term plan. The active mission phase ended in April 2013 after depletion of the on-board helium cryogen supply. As a space-based observatory, \textit{Herschel} provided an unobstructed view of the far-infrared (FIR) Universe in orbit about the L2 Lagrangian point of the Earth/Sun system. The observatory was equipped with a heat shield, allowing passive cooling of the telescope to $\sim $85\,K, and a 3.5\,m diameter primary mirror offering unprecedented angular resolution in the FIR. The payload module housed three science instruments: the Spectral and Photometric Imaging Receiver (SPIRE; \cite{Griffin10}), the Photodetector Array Camera and Spectrometer (PACS; \cite{poglitsch2010photodetector}), and the Heterodyne Instrument for the Far-Infrared (HIFI; \cite{de2010herschel}). A multi-stage cryostat cooled these instruments to $ \sim $1.7\,K with a dedicated closed-cycle  $ ^{3} $He sorption cooler providing an optimal $ \sim $300\,mK environment for the SPIRE and PACS bolometer detectors. \emph{Herschel's} instruments covered a spectral range from 55 to 676\,$ \mu $m, corresponding to the peak intensities of black bodies between 5 to 50\,K, and a photon energy of 1.8 to 22.5\,meV. Thus \textit{Herschel} is sensitive to the thermal emission of cold dust grains, and low energy rotational and fine structure emission. This makes \emph{Herschel} ideal for observing the icy outer solar system, cold molecular clouds in the interstellar medium (ISM), and red shifted galaxies. 

The SPIRE instrument is comprised of an imaging Fourier Transform Spectrometer (FTS) and a three-band imaging photometer. The FTS detector arrays consisted of two hexagonally close-packed bolometer detector arrays: the SPIRE Short Wavelength array (SSW; 959.3-1544\,GHz) and the SPIRE Long Wavelength array (SLW; 446.7-989.4\,GHz), providing simultaneous spectral imaging over a broad spectral range. All science observations were carried out in either high resolution (HR), or low resolution (LR) mode corresponding to spectral resolutions of $ \Delta f = 1.2 $\,GHz and $ \Delta f = 25.0 $\,GHz, respectively \citep{spire_handbook}.

A wealth of FTS data was collected by SPIRE during \textit{Herschel's} mission including 1850 FTS science observations. Analysis of FTS data can be a challenging and time consuming process due to the complicated instrument line shape, which can be approximated as a sinc function. To this end, the SPIRE FTS Spectral Feature Finder \cite[FF; \citetalias{FFtech}:][]{FFtech} has been developed to assist with the preliminary analysis of SPIRE FTS data and expedite data-mining by providing a publicly available spectral feature catalogue. This tool was developed for convenience and is not intended to replace detailed analysis. The FF routine, validation, and products are described fully in the companion paper \citetalias{FFtech}, with only a general summary presented below. 

\begin{figure}
\centering
\includegraphics[width=1.0\linewidth]{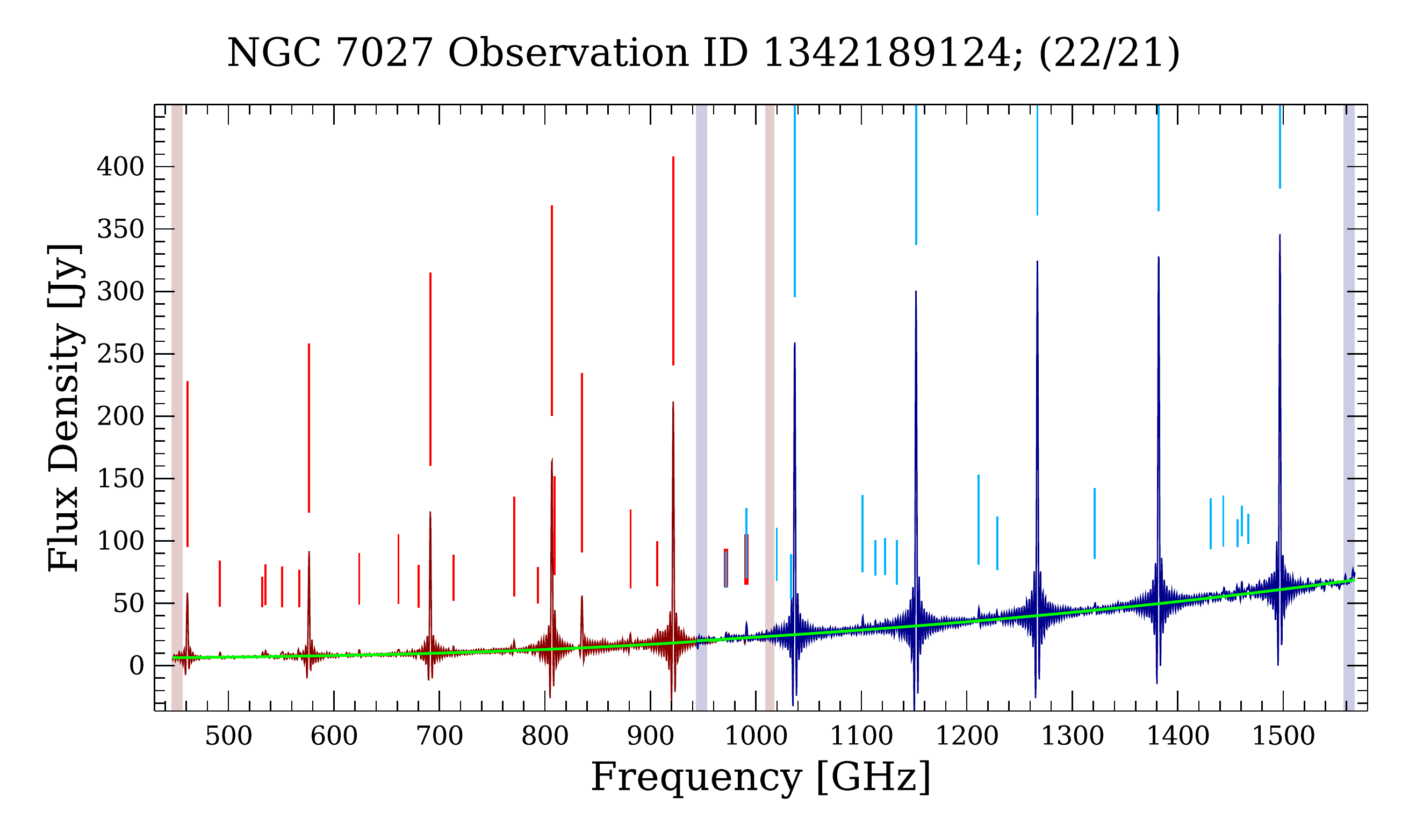}
\vspace{-18pt}
\caption{A sample postcard from the FF sparse catalogue. NGC 7027 is a young planetary nebula showing strong rotational \CO emission. SPIRE SLW and SSW spectral data are shown with red and blue curves, respectively. The fitted continuum is indicated with a green curve, and detected spectral features are indicated with vertical lines. The height of each vertical line is proportional to the log of the absolute value of the SNR of the corresponding spectral feature. Red and blue vertical bars extending the height of the plot highlight the noisy band edge regions of the SLW and SSW spectra, respectively. These regions were ignored by the FF routine.}
\label{fig:1342189124_postcard1}
\end{figure}

\begin{figure*}
\centering
\includegraphics[width=1.0\linewidth]{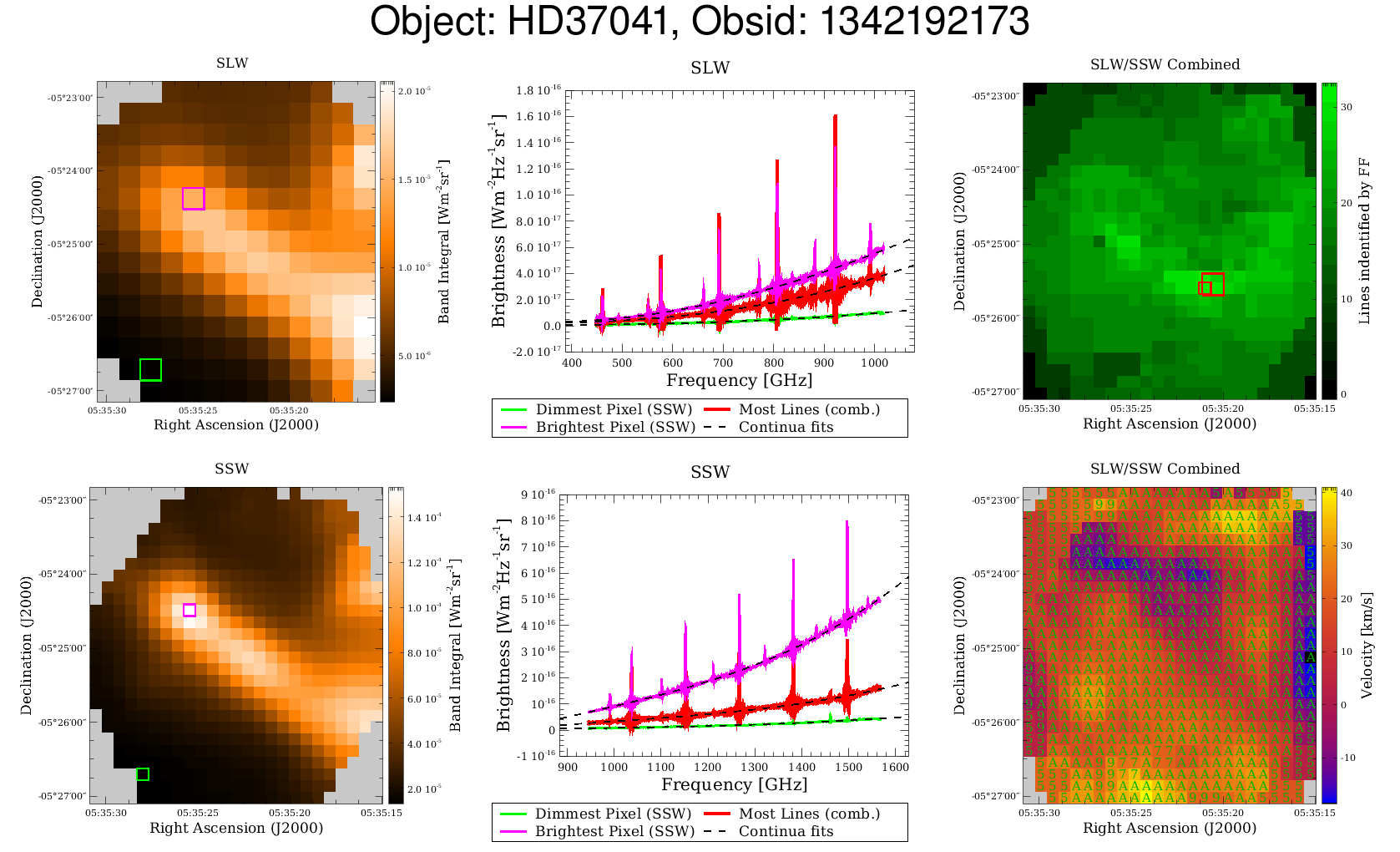}
\vspace{-12pt}
\caption{A sample postcard from the FF mapping catalogue. The source under observation is the Orion bar. Integrated flux maps for SLW (top-left) and SSW (bottom-left) spectral images are presented in the first column. These maps contain magenta and green squares indicating the brightest and faintest pixels from overlapping SLW/SSW regions, respectively. Each pixel contains a spectrum which is processed by the FF. The number of detected features within the SLW and SSW maps are combined into a single map (see \S\,\ref{sec:MethMapping}), with the total number of features within the resulting map shown in the top right panel. Red squares show the co-added pixels with the most detected features. The spectra corresponding to the aforementioned highlighted pixels are plotted in the centre column using a consistent colour coding. SLW (top-centre) and SSW (bottom-centre) spectra are kept separate for clarity, with the fitted continua plotted using black dashed curves. A velocity map, calculated using the routine outlined in this paper, is shown in the bottom right panel. Annotations in the velocity map correspond to the velocity estimate's associated flagging parameter which is discussed in \S\,\ref{sec:MethMapping}.}
\label{fig:1342192173_postcard_comb}
\end{figure*}

The FF accommodates both sparse and mapping observations, and operates on unapodized HR and LR SPIRE spectra. LR spectra are generally not of sufficient spectral resolution to identify individual spectral features, therefore, for these observations only the coefficients of the second order polynomials used to model the continua are presented. For HR spectra, frequencies and associated signal-to-noise ratios (SNR) of significant spectral emission/absorption features are presented in addition to the coefficients of the third order polynomials used to model the continua. A FF flag accompanies each spectral feature, which acts an indicator of the feature's detection quality. Integrated flux is not included in part because of the complications associated with partially resolved spectral features, but also to encourage follow-up analysis from users. Accompanying each observation is a postcard giving a visual representation of the FF results, with each HR and LR observation receiving their own postcards. An example of an HR sparse FF postcard is shown in Fig.\,\ref{fig:1342189124_postcard1}, and an HR mapping postcard is shown in Fig.\,\ref{fig:1342192173_postcard_comb}. The FF products are compiled into the publicly available SPIRE Automated Feature Extraction CATalogue (SAFECAT\footnote{https://www.cosmos.esa.int/web/herschel/spire-spectral-feature-catalogue}), as part of the Herschel Science Archive (HSA). 

Feature detection within the FF is accomplished using an iterative approach. A spectral model is constructed using a third order polynomial for the continuum, and a sinc function for each spectral emission or absorption feature. Continuum parameters are first estimated by masking strong features, identified using a differential measurement between LR and HR spectra from the same observation, then fitting the unmasked regions. Potential spectral features are initially identified by searching for peaks above a given threshold in the observation's SNR spectrum. The SNR spectrum is calculated as the spectral flux density divided by the corresponding spectral uncertainty. Initial frequency and amplitude estimates for each sinc function are derived from the frequency of the SNR peak, and the continuum subtracted flux density of the spectral data at that frequency, respectively. The majority of SPIRE spectral features are unresolved, thus the width of sinc models are fixed to a constant value $\Delta f/\pi~\sim$0.38\,GHz. A global fit initiates and features meeting a rigorous selection criteria are added to the total model of the spectrum. Continuum parameters resulting from this fit become the initial estimate for the continuum in the next iteration. Accepted features are subtracted from the spectrum, and the above process is repeated with a lower SNR threshold. After the routine has iterated over all SNR thresholds a spectral residual is calculated as the difference between the final spectral model and the spectral data. For each feature, the final SNR is calculated as the ratio of the fitted feature's amplitude and the root mean square of the residual from the spectral region surrounding the detected feature. 

The FF can reliably detect strong spectral features. Most importantly for this paper, the FF shows a $ \sim $98\% detection rate for lines with an SNR $ \ge 10 $ within a large dataset of 20,000 simulated spectra containing the $^{12}$CO ladder, and four other random features \citepalias{FFtech}. Thus, the reliable detection of significant $^{12}$CO features is well supported. 

In addition to the continuum and features themselves, another parameter of interest is the source radial velocity. Radial velocity is useful for characterizing astronomical sources by probing the dynamics of the ISM, identifying the signatures of accretion disks and bipolar outflows in protostars, and is proportional to distance for extragalactic sources. As such, an attempt is made to estimate the source radial velocity in the products of the FF. This is accomplished using two dedicated subroutines. The first, discussed in \S\,\ref{sec:CORoutine}, performs a search for rotational \CO emission and pairs these identified features to their nearest rest frequency. The second, discussed in \S\,\ref{sec:xcor_routine}, performs a cross-correlation between detected features and a line template consisting of molecular and atomic lines commonly detected in the FIR. For observations where no reliable velocity estimate is obtained using these methods, we consult external references as outlined in \S\,\ref{sec:externalRef}. The results of these routines as they pertain to the FF are summarized in \S\,\ref{sec:FF_Results}. The limitations of our routines are discussed in \S\,\ref{sec:discussion}. Summaries of our velocity estimating methods and concluding remarks are presented in \S\,\ref{sec:Conclusion}.

This paper is one of a series of four which discuss aspects of the FF: the main technical paper \citep[\citetalias{FFtech}:][]{FFtech}, this radial source velocity paper, a line identification paper \citep[\citetalias{FFlineID}:][]{FFlineID} which also presents FF results for the off-axis spectra within sparse observations, and a [CI] detection and deblending paper \citep[\citetalias{FFncc}:][]{FFncc}.

\section{\texorpdfstring{\CO}{12CO} Identification Routine}
\label{sec:CORoutine}

The source radial velocity, $ \mathrm{v} $, is related to rest frequency, $ f_0 $, measured frequency, $ f $, and the speed of light, $ c $, by the equation,
\begin{equation}
\label{eq:doppler}
\mathrm{v} \approx \left(\frac{f_0}{f} - 1 \right)  c = zc \;.
\end{equation}
This is the conventional definition consistent with the non-relative approximation $ v \approx zc $, with $ z $ corresponding to the astronomical redshift. Eq.\,\ref{eq:doppler} produces what is sometimes referred to as the optical velocity, and as written, a positive velocity is achieved for sources moving away from observer. This approximation is accurate only for low redshifts. Although the redshifts obtained from the routines described in this paper have $|z| < 1.0$, the estimates obtained from external references (see \S\,\ref{sec:externalRef}) do exceed the range of validity. We maintain the approximation regardless for consistency and for the simplicity of converting observed frequencies of detected spectral features to their rest frequencies. 

The most direct way of estimating the source radial velocity is to identify a line within the source's spectrum and match it with its corresponding rest frequency, and then employ Eq.\,\ref{eq:doppler} to determine $\mathrm{v}$. The routine outlined in this section applies the above procedure to \CO rotational emission features within the FF detected spectral feature list for a given SPIRE observation. 

$ ^{12} $CO is the second most abundant molecule in the Universe next to H$ _2 $ \citep{herbst1995chemistry}. Being a symmetric diatomic molecule, H$_2$ possesses no electric dipole, and only a weak quadrupole resulting in very weak H$_2$ ro-vibrational emission within cold molecular clouds. On the other hand, \CO has a dipole resulting in a bright rotational emission spectrum in the FIR. In the rest frame, \CO expresses a total of 10 rotational emission features in the SPIRE bands (\mbox{J=4-3} to \mbox{J=13-12}, see Fig.\,\ref{fig:1342189124_postcard1}), which is often referred to as the $ ^{12} $CO ladder. Thus, \CO is an ideal candidate for a generalised approach to determine source radial velocity; the lines are not only likely to be present, but also reliably identified with an automated routine. 

\begin{figure}
\centering
\includegraphics{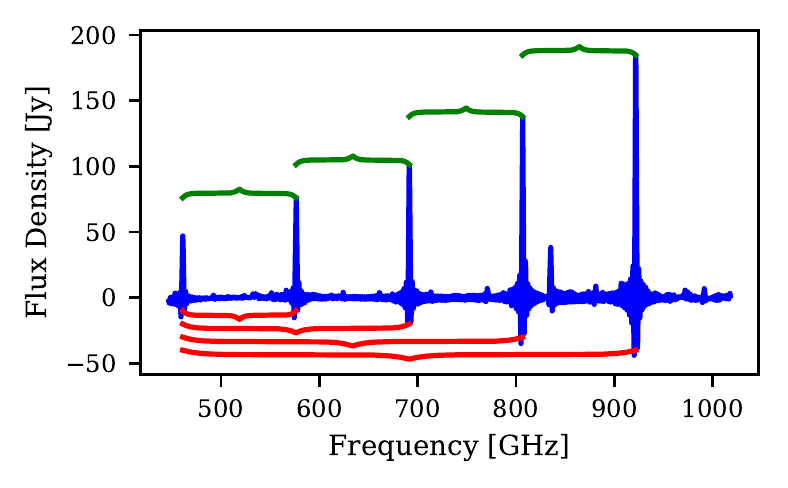}
\vspace{-10pt}
\caption{Continuum subtracted SLW spectrum of NGC 7027. Upper braces indicate the near constant frequency difference between adjacent \CO rotational emission features. The second to fifth entries of the CDA are illustrated by the lower braces with the first entry omitted due to its zero width.}
\label{fig:Spectrum1}
\end{figure}

\begin{figure}
\centering
\includegraphics{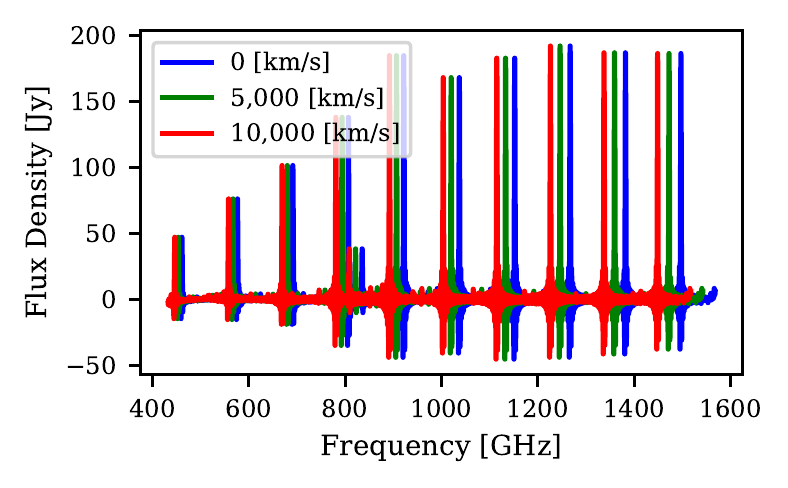}
\vspace{-12pt}
\caption{The continuum subtracted spectrum for NGC 7027 at various redshifts. Notice the frequency difference between adjacent emission features, and thus the CDA, changes with source velocity.}
\label{fig:redshiftedSpectrum}
\end{figure}

Using the rigid rotor model for diatomic molecules, the frequency difference between adjacent rotational emission features is given by 
\begin{equation}
\Delta f = \frac{h}{4\pi^2I} \;,
\label{eq:rigidRotator}
\end{equation}
where $ h $ is Planck's constant, and $ I $ is the moment of inertia of the molecule \citep{hollas2004modern}. Thus, to first order, the frequency spacing between diatomic rotational emission lines is a constant value determined by the mass distribution of the molecule. For \CO this value is $ \sim $115.1\,GHz and remains fairly constant over the SPIRE band, varying by only $ \sim $0.2\,GHz after accounting for centrifugal distortion for higher energy transitions. Keeping in mind the spectral resolution of SPIRE in high resolution mode is 1.2\,GHz, a 0.2\,GHz deviation is not significant in our search algorithm. Our method of searching for \CO rotational emission features is based on searching frequency differences between detected lines for the characteristic frequency differences of the \CO rotational emission ladder ($ \Delta f_n \approx n\times 115.1 \, \mathrm{[GHz]},  \,\,\   n = 0,1,2,...,9$). We refer to this sequence of vales as the Characteristic Difference Array (CDA) for $ ^{12} $CO. Note that zero is included in the CDA to simplify the detection algorithm, and corresponds to the frequency difference between a \CO feature and itself. Fig.\,\ref{fig:Spectrum1} shows the SLW spectrum of an NGC 7027 observation and helps to illustrate these concepts. The upper braces represent the near constant frequency difference between adjacent $ ^{12} $CO rotational emission features, with the lower braces representing the second to fifth entries of the CDA (the first entry not shown due to its zero width). However, the frequency difference between adjacent $ ^{12} $CO features is not constant for sources with a non-zero radial velocity. Fig.\,\ref{fig:redshiftedSpectrum} shows the continuum subtracted spectrum for NGC 7027 with various induced redshifts. It is evident that the CDA is no longer preserved with the $ ^{12} $CO ladder being somewhat compressed towards lower frequencies at higher redshifts. Accounting for this effect, the CDA is modified to

\vspace{-12pt}
\begin{equation}
\mathrm{CDA'} = \left(\frac{\mathrm{v}}{c} + 1 \right)^{-1} \times \mathrm{CDA} \;.
\label{eq:difference}
\end{equation}
The difference between the CDA and the velocity dependent version is given by,
\vspace{-6pt}
\begin{equation}
\label{eq:tolerance}
\mathrm{tol} = \mathrm{CDA' - CDA} = \left( 1 + \frac{c}{\mathrm{v_\mathrm{max}}} \right)^{-1} \times \mathrm{CDA} \;.
\end{equation}
If we intend to search the frequency differences between detected lines for elements matching the CDA, Eq.\,\ref{eq:tolerance} can be interpreted as a velocity dependent tolerance on the agreement between the CDA and measured frequency differences. The velocity in Eq.\,\ref{eq:tolerance} then takes the role of some maximum assumed velocity. In practice, this parameter need only be larger than the radial velocity of the source to obtain good results. Eq.\,\ref{eq:tolerance} expresses an asymmetry between positive and negative velocities, but the difference is only about 10\% at the limits of the viability of this routine ($ \pm 14,000 $\,\kms, \S\,\ref{subsec:methodCO}). As such, the tolerance calculated for a velocity of a given magnitude is sufficiently applicable to both positive and negative radial velocities. 

\begin{figure}
\centering
\includegraphics[width=1.0\linewidth]{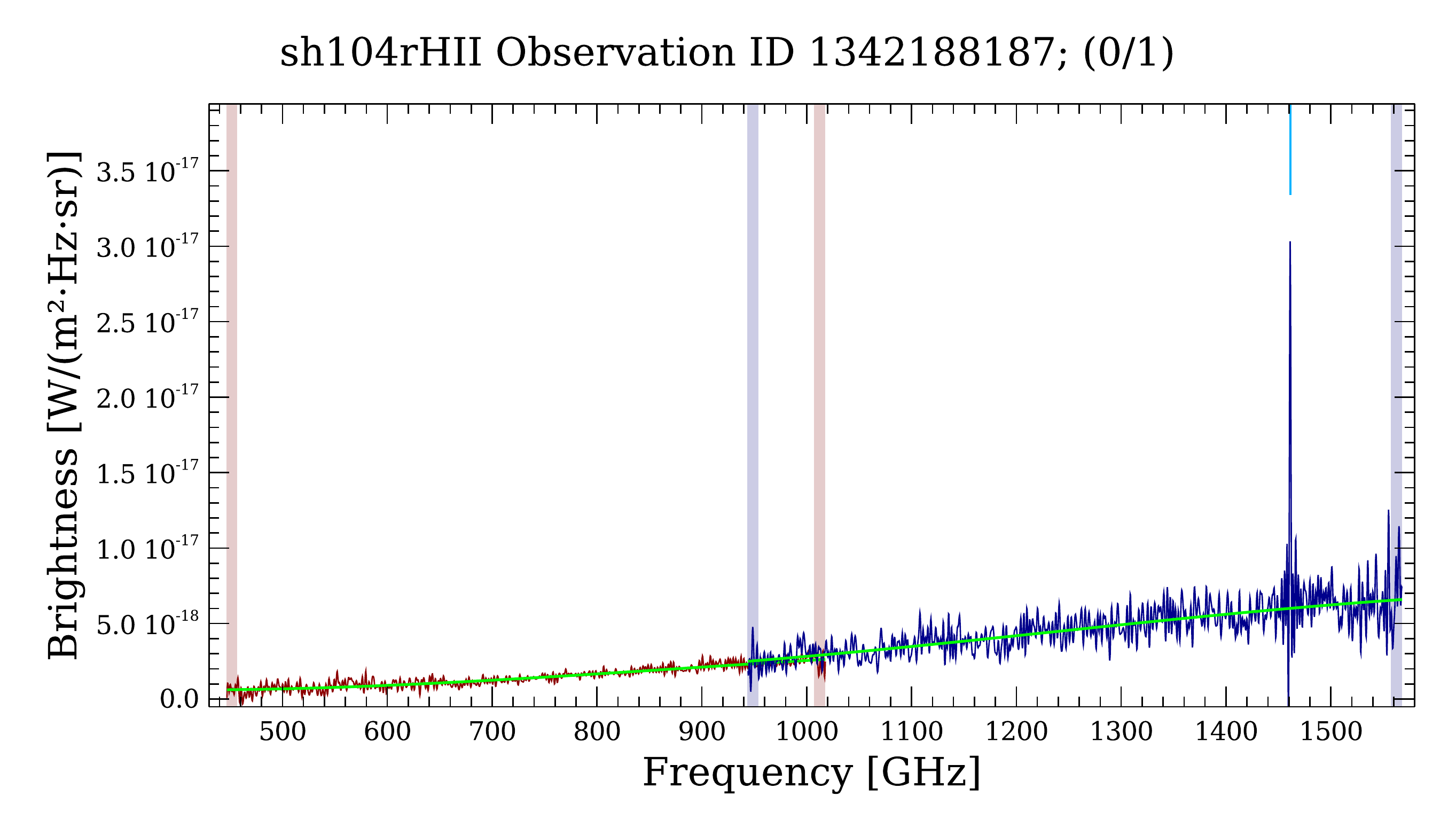}
\vspace{-18pt}
\caption{FF postcard for the star forming region sh-104. The only line evident in the spectral data is the [NII] feature.}
\label{fig:1342188187_postcard}
\end{figure}

\begin{figure}
\centering
\vspace{-9pt}
\includegraphics[width=1.0\linewidth]{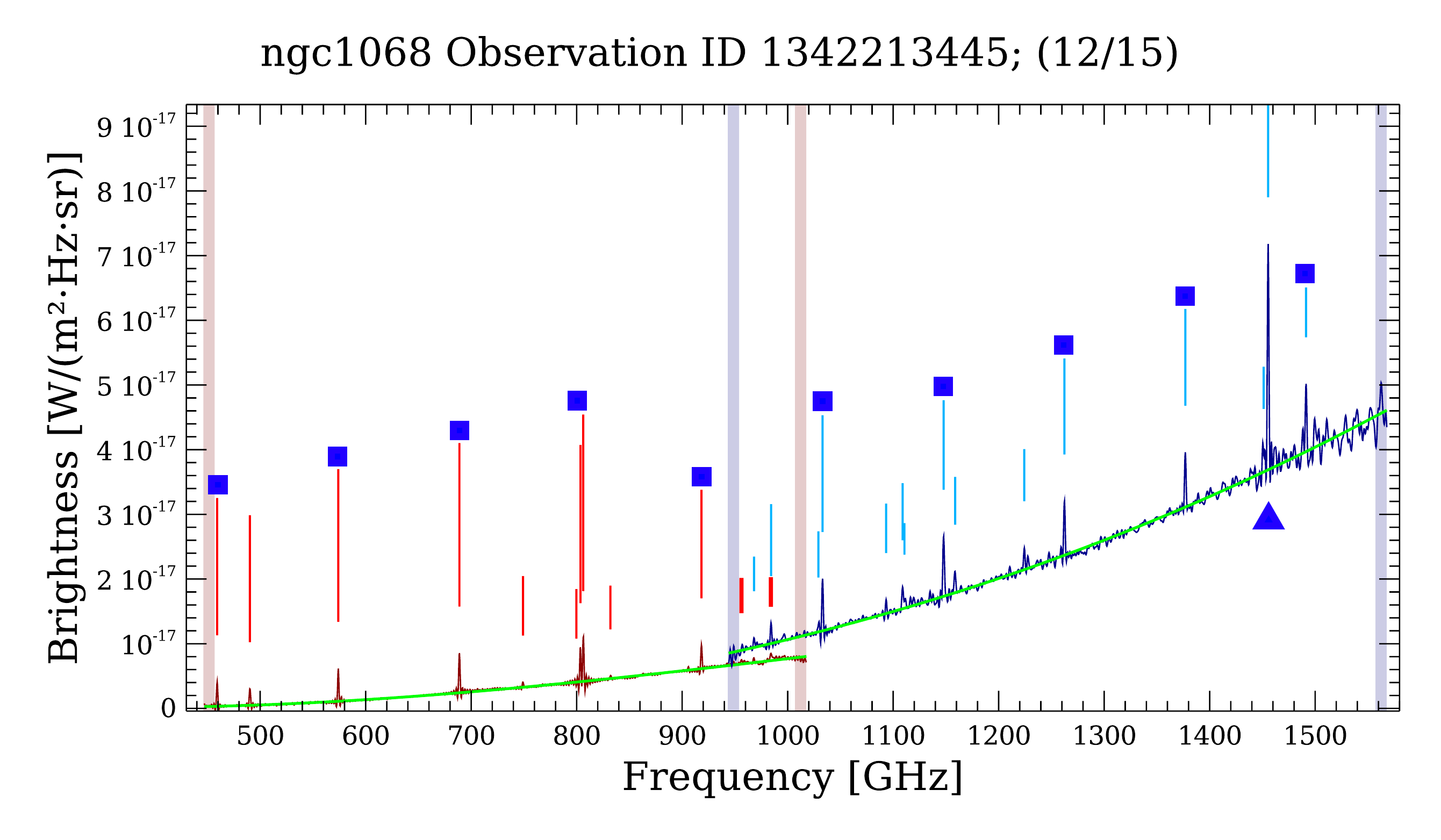}
\vspace{-12pt}
\caption{An example of a source containing both \CO rotational emission and [NII] emission in the same spectrum. \CO emission is indicated with blue squares, and [NII] emission is indicated by the blue triangle.}
\label{fig:NIIandCOinView}
\end{figure}

Some sources exist, like the massive star forming region sh-104 (Fig.\,\ref{fig:1342188187_postcard}), which exhibit no identifiable \CO rotational features, but contain a prominent ionized nitrogen fine structure line ([NII]$\mathrm{^3P_1 - ^3P_0}$; 1461.13\,GHz). These spectra occur since rotational \CO and [NII] probe separate energy regimes of the interstellar medium. Nitrogen has an ionization energy of $\sim$14.53\,eV \citep{zhao2016n}, while \CO dissociates at $\sim$11.11\,eV \citep{pauling1949dissociation}, and so it is not atypical to see solitary [NII] features in spectral data. This phenomenon was kept in mind while developing the FF radial velocity estimating routine. \CO and [NII] can also be be observed within the same spectrum, indicative of multiple energy regimes within the same field of view. An example of this phenomenon is NGC 1068, a barred spiral galaxy expressing a foreground ionization component and a background galactic disk containing molecular gas. Fig.\,\ref{fig:NIIandCOinView} shows the spectrum of NGC 1068 with blue squares to indicated \CO rotational emission, and a blue triangle indicating the strong [NII] emission.

\subsection{Method}
\label{subsec:methodCO}

The \CO based radial velocity estimating routine takes as input a list of frequencies, frequency errors, and SNR values corresponding to emission features (i.e. those with positive SNR) detected by the FF. For each line, we calculate the frequency difference array (FDA) between it and all lines in the list (including itself), resulting in a difference array for each input line. Each FDA is then searched for elements that match the CDA. To accomplish this, we iterate through the CDA and identify any elements in the difference array that match to within the velocity dependent tolerance (Eq.\,\ref{eq:tolerance}). If there is at least one element that meets this criteria, a `match' counter is incremented by one. In general, difference arrays can have negative values, so a similar process occurs with a negated version of the FDAs. This process is then repeated for all FDAs (i.e. for every spectral emission feature detected by the FF).

The lines with the greatest number of matches have the greatest probability of being \CO\!. These lines are extracted and paired to the \CO transition with the nearest rest frequency. This step imposes a fundamental limit on the highest velocity magnitude that can be estimated reliably. As can be seen from Fig.\,\ref{fig:redshiftedSpectrum}, the highest frequency \CO line (J=13-12) approaches the J=12-11 line with increasing source velocity. At $ \sim $12,000\,\kms, the shifted J=13-12 line is in closer spectral proximity to the rest frequency of the J=12-11 line than its own rest frequency, so it is erroneously paired. As will be described shortly, the routine is somewhat resilient to this effect and such erroneously paired \CO candidates are generally discarded. The threshold for a good velocity estimate is one which finds at least 7 \CO candidate features, and as such the routine is expected to break for velocities in excess of $ \sim $14,000\,\kms. 

Due to the often large tolerance parameter, it is possible that multiple lines will be attributed to the same \CO transition if the source has a high spectral line density. In these cases, the line with the highest SNR is kept, and the others are discarded. A smaller tolerance can help reduce this effect, so it is preferential to use a small maximum velocity resulting in a smaller tolerance when performing this check. Therefore, the tolerance is first calculated with a maximum velocity of 6,000\,\kms. If the routine does not converge on a reliable estimate, the maximum velocity is then increased by increments of 2,000\,\kms, up to a maximum value of 14,000\,\kms. The total number of potential \CO lines found can be used as a measure of confidence on the velocity estimate. For this reason, the total number of candidate lines used for calculating a velocity estimate is recorded and assigned to the parameter `N'.  

Once the \CO candidate features have been paired with their corresponding rest frequencies, a velocity is calculated for each line using Eq.\,\ref{eq:doppler}. A further check is performed on the quality of the estimates by calculating the standard deviation of these velocities. A standard deviation greater than 100\,\kms\ indicates unreliable \CO detection as all \CO lines are expected to have a similar redshift. When this occurs the line with the highest deviation from the median velocity is removed. This process is repeated until the standard deviation of velocities is below 100\,\kms. Attempting to reduce the standard deviation of the constituent velocities in this way is effective at removing \CO candidate features erroneously paired to the wrong rest frequency.

As previously mentioned, SPIRE spectra exist which contain prominent [NII] fine structure emission with no detectable \CO rotational emission. A simple exception is included in our routine to account for such occurrences. If no \CO is found, and the FF detected 10 or fewer features, we search 60\,GHz on either side of the rest frequency of [NII]. The feature with the highest SNR in this window is assumed to be [NII], and if the corresponding feature's SNR is $ \ge $10, a velocity estimate is calculated on this assumption. Note that [NII] probes a different energy regime than $ ^{12} $CO. Thus, the calculated source velocity may vary dependent on which species is used. For example, velocity calculations for NGC 1068 (Fig.\,\ref{fig:NIIandCOinView}) based on [NII] and \CO emission return a $ \sim $61\,\kms\ difference.

Validation of this routine is provided in \S\,\ref{subsec:Validation_CO}. Here we outline the FF criteria for a `high quality' radial velocity estimate. Estimates based on \CO candidate features with fewer than two matches with the CDA were discarded. In the sparse observation dataset, velocity estimates with \mbox{N $ > 6 $}, or based on an [NII] feature were considered sufficiently reliable and incorporated into the final data products of the FF. For mapping observations, SLW and SSW spectral images generally do not occupy the same footprint on-sky due to their beam size disparity. As a result, the peripheries of the combined images (see \S\,\ref{sec:MethMapping}) often contain only SLW spectra which nominally express a maximum of 5 \CO rotational emission features. For this reason the FF acceptance criteria for mapping observations was reduced to estimates with \mbox{N $ > 3 $} if the pixel contains only features from the SLW band. Mapping observations with velocity estimates based on [NII] emission were also considered reliable.

The velocity error is calculated in one of two ways depending on how many features are used to generate the velocity estimate. If multiple spectral features are used, the error is calculated as the standard deviation between the velocity estimates corresponding to each candidate feature. If only one feature is used, as is the case for estimates based on the [NII] feature, we propagate the error associated with the line centre frequency, determined during the fitting process, into velocity space.

\subsection{Mapping Observations}
\label{sec:MethMapping}

In order to maintain consistency with the sparse catalogue, and to preserved the confidence that comes with greater quantity of \CO emission feature detection, we attempt to combine SLW and SSW line lists for mapping observations. Due to their different beam sizes, the SLW and SSW bands vary in both sky coverage and pixel dimensions. Combining the line lists from the two spectral images is accomplished by projecting the SLW lines onto an SSW equivalent grid. This is achieved by simply combining the line list from a SSW pixel with the line list of the nearest SLW pixel. This process can result in a small amount of spatial smearing since the combined pixels do not always align on-sky, however, the centres of the two pixels are limited to a maximum separation of an SLW pixel width divided by $\sqrt{2}$, so the smearing is expected to be small. 

The nominal SLW beam size is $\sim$1.8 times larger than the nominal SSW beam resulting in greater sky coverage in the SLW maps compared to the SSW maps. As such, The perimeter of the combined maps tend to have spectral features from only the SLW band. As previously mentioned, since the SLW band nominally expresses only 5 \CO  rotational emission features, the velocity estimate acceptance criteria has been reduced to N $>$ 3 for pixels containing only SLW features. 

Once the SLW features have been projected onto the SSW grid, the radial velocity estimating routine operates on each pixel producing a radial velocity map. An example of one of these maps is shown in the bottom right corner of Fig.\,\ref{fig:1342192173_postcard_comb}. Each pixel in the velocity map contains an annotation indicating the N flagging parameter associated with each estimate. In the case where 10 \CO candidate features were used to generate the velocity estimate, the annotation is `A' indicating all \CO features were identified. When [NII] emission is used, the annotation is `N'. 

\subsection{Validation}
\label{subsec:Validation_CO}

Validation of the \CO based radial velocity estimating routine is accomplished using comparisons with simulated (\S\,\ref{subsubsec:SimData}) and real (\S\,\ref{subsubsec:RealData}) data. Comparisons with real data consists of velocity estimates derived through HIFI observations and manual inspection of SPIRE spectral data. 

\subsubsection{Comparison with simulated data}
\label{subsubsec:SimData}
To test the reliability of the radial velocity estimating routine over a broad range of velocities, we simulated FF products for 20,000 realistic spectra. The output for each simulated spectrum was a list of spectral emission frequencies, with corresponding SNRs and frequency errors. We begin generating the line list by introducing the $^{12}$CO ladder rest frequencies with their corresponding SNR values derived from the FF data products for the co-added NGC 7027 data \citep{Hopwood15}. The $^{12}$CO features in NGC 7027 are of high SNR with little line blending compared to the other spectral calibration sources \citep{Hopwood15}. The strong $^{12}$CO features makes NGC 7027 a good candidate simulation template, while providing a soft upper limit on the spectral line SNR values. The $^{12}$CO SNR values are then individually scaled by a factor between 0.1 and 1.1. There was then a 25\% probability that between 1 and 4 $^{12}$CO lines would be randomly removed. A frequency shift was induced by randomly selecting a radial velocity between -15,000 and +15,000\,\kms. The $^{12}$CO SNR scaling factor, number of lines removed, and radial velocity were selected from a random uniform distribution. We then introduced between 10 and 25 additional non-$^{12}$CO features with frequencies between 447 and 1546\,GHz. Their corresponding SNRs were limited such that the maximum value would be 105, with the minimum limit set to 5 in agreement with the minimum FF detection threshold. The number of lines, frequency positions, and SNR values were again chosen from a uniform random distribution. Both the $^{12}$CO and extra features were concatenated into a single list of features. We induce an additional layer of uncertainty by varying the frequencies of each line using a random normal distribution with sigma equal to the mean line centre error reported by the FF for the NGC 7027 co-added data (0.11\,GHz). The SNR values are likewise concatenated into a single list, with the frequency error list being a constant 0.11\,GHz for all features.

\begin{figure}
\centering
\vspace{-6pt}
\includegraphics{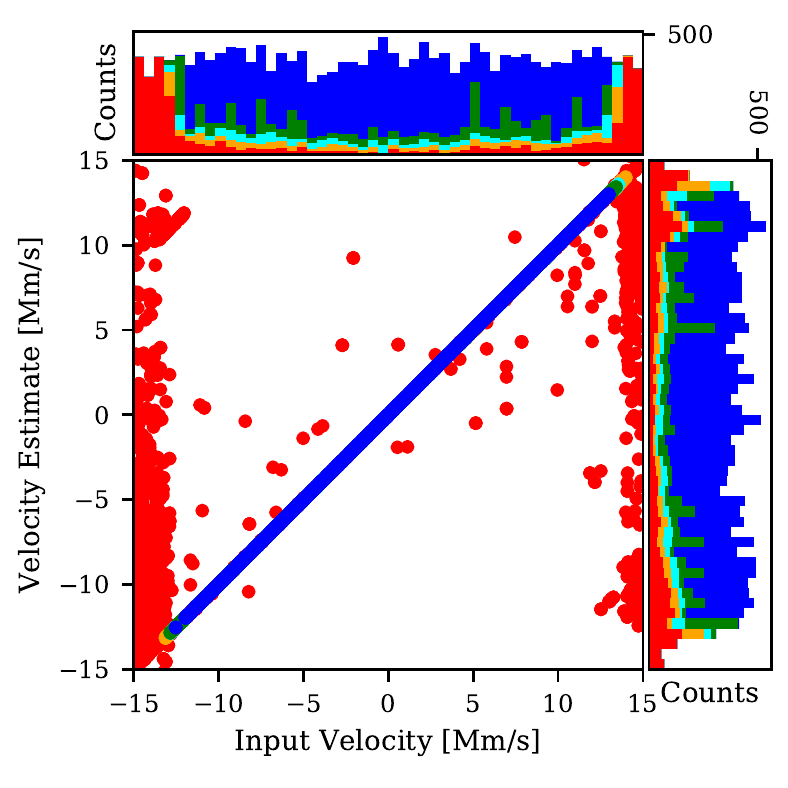}
\vspace{-15pt}
\caption{Velocity estimate correlation plot for the simulated dataset. Colour-coding is based on the N flagging parameters with (10--9: blue, 8--7: green, 6--5: cyan, 4--3: orange, 2--1: red). No high quality estimates (N > 6) deviate significantly from the one-to-one line. Histograms on top and to the right of the central figure show the distribution of input and estimated velocities, respectively.}
\label{fig:simCorr}
\end{figure}

The simulation products were processed by the radial velocity estimating routine. A correlation plot of the results is shown in Fig.\,\ref{fig:simCorr}. It can be seen that estimates with an N $>$ 3 match the input velocity quite well. There is an immediate breakdown in reliability of the routine beyond a velocity magnitude greater than 14,000\,\kms\ as was expected. This breakdown is marked by a reduction in the value of the flagging parameter, also as expected. We note that there are no velocity estimates with a flagging value greater than 6 which deviate significantly from the input velocity. The absolute value of the residual velocities for each flagging parameter are shown in Fig.\,\ref{fig:flagResidual} with blue markers. The residual velocities are on the order of what would be expected from the uncertainties implemented in the simulated data. A completeness curve of the residual velocities with an inset showing their distribution is presented in Fig.\,\ref{fig:Completeness}, again using blue. Only estimates with \mbox{N $ > 3 $} were used in Figs.\,\ref{fig:flagResidual} and \ref{fig:Completeness}, in agreement with the minimum flagging criteria for mapping observation velocity estimates. It can be seen that $\sim$90\% of estimates have a residual velocity $\le$ 20\,\kms. Isolating all estimates with a residual velocity $ \le $\,20\,\kms\ and residing within an input velocity range of $ \pm $14,000\,\kms, we find that 84\% of estimates have \mbox{N $ > 6 $}. That is, our selection criteria captures about 84\% of accurate velocity estimates, while including no inaccurate estimates. 

\subsubsection{Comparison with real data}
\label{subsubsec:RealData}

\begin{figure}
\centering
\vspace{-6pt}
\includegraphics{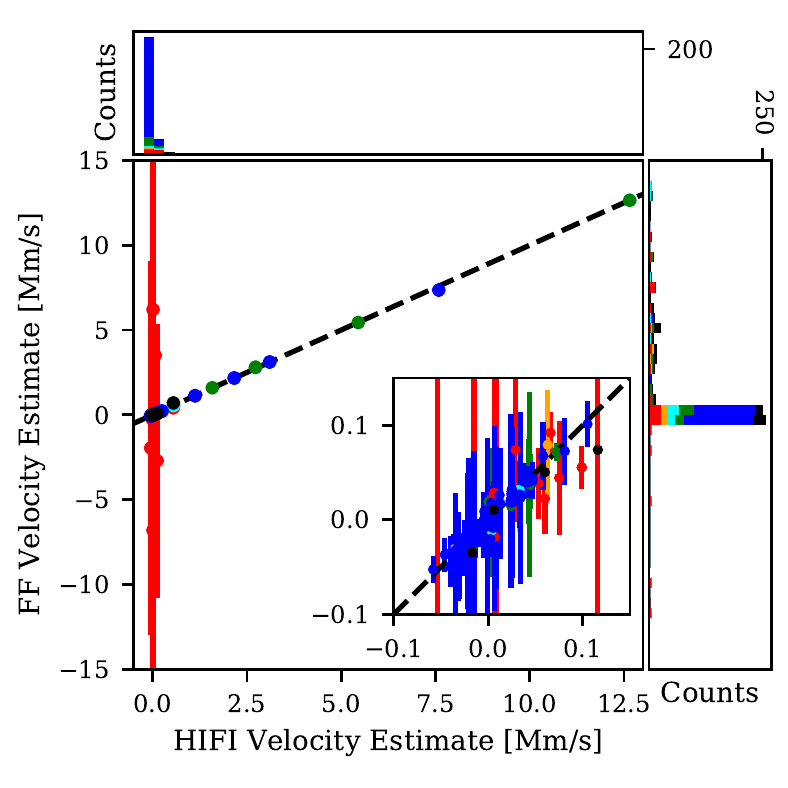}
\vspace{-15pt}
\caption{Velocity estimate correlation plot for the complimentary HIFI dataset using the same colour scheme as in Fig.\,\ref{fig:simCorr}. Estimates derived from [NII] emission are indicated with black markers. Inset shows a close-up of the lower velocity cluster. No high quality estimates deviate significantly from the one-to-one line. Histograms on top and to the right of the central figure show the distribution of HIFI and estimated velocities, respectively.}
\label{fig:hifiCorr}
\end{figure}

\begin{figure}
\centering
\vspace{-3pt}
\includegraphics{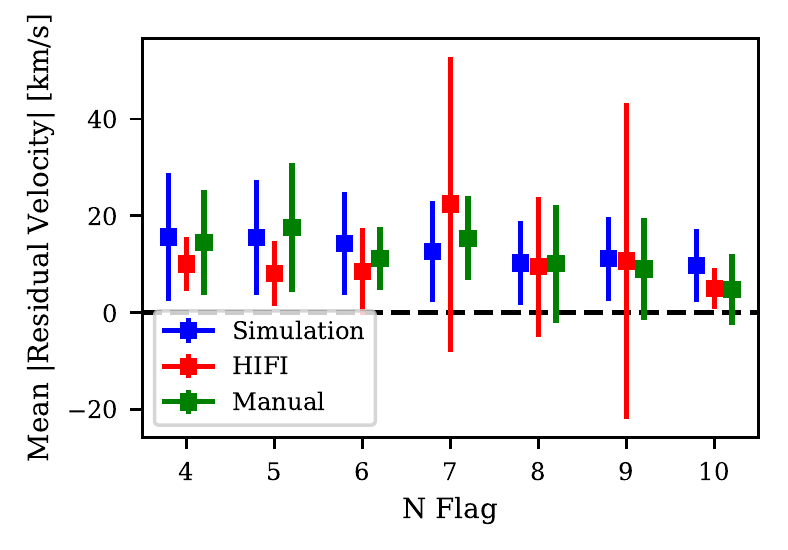}
\vspace{-15pt}
\caption{Mean absolute residual velocity of velocity estimates derived from the \CO based velocity estimating routine and the velocity estimate derived from three comparison datasets. The distribution is given in terms of the N flagging parameter and shows only estimates with \mbox{N $ > 3 $}. Blue, red, and green markers indicated results obtained from the simulation, HIFI, and manual inspection comparison datasets, respectively.}
\label{fig:flagResidual}
\end{figure}

\begin{figure}
\centering
\includegraphics{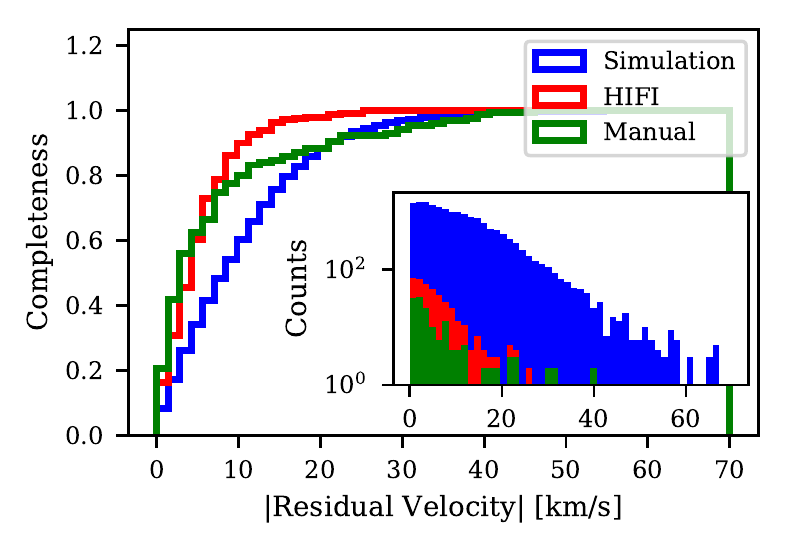}
\vspace{-15pt}
\caption{Completeness curves of the absolute value of the residual velocities for the three comparison datasets with inset showing the distribution of these residuals. For the simulation and manual inspection datasets, shown in blue and green, respectively, $ \sim $90\% of residual velocities are $ \le $20\,\kms. For the HIFI dataset, shown in red, $ \sim $90\% of residual velocities are $ \le $10\,\kms. The figure includes only velocity estimates with an N flagging parameter greater than 3.}
\label{fig:Completeness}
\end{figure}

The first real dataset consists of radial velocity estimates derived from HIFI \citep{hifiVelocity}, which form a complimentary dataset to a subset of SPIRE observations. A correlation plot of complimentary HIFI and FF radial velocity estimates is shown in Fig.\,\ref{fig:hifiCorr}. The average absolute value of the residual velocities in terms of the N flagging parameter is shown in Fig.\,\ref{fig:flagResidual} using red markers. The completeness curve for the absolute value of the residual velocities is shown in Fig.\,\ref{fig:Completeness}, again using red, with an inset showing the residual velocity distribution. The completeness curve and corresponding histogram use only estimates with \mbox{N $ > 3 $}.

\begin{figure}
\centering
\vspace{-6pt}
\includegraphics{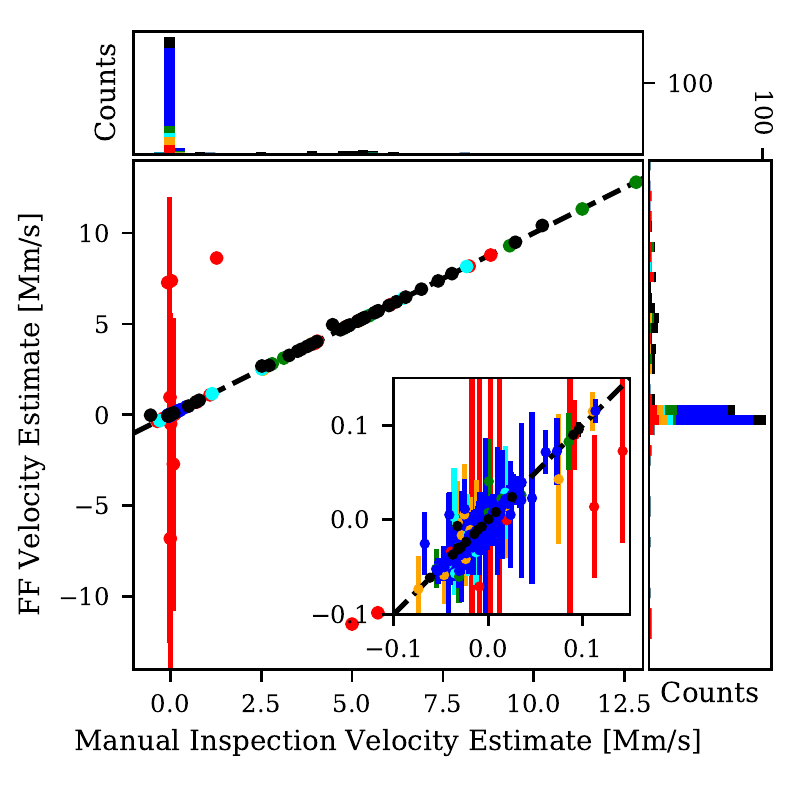}
\vspace{-15pt}
\caption{Velocity estimate correlation plot for the manual inspection dataset using the same colour scheme as in Fig.\,\ref{fig:simCorr}. Estimates derived from [NII] emission are indicated with black markers. Inset shows a close-up of the lower velocity cluster. No high quality estimates deviate significantly from the one-to-one line. Histograms on top and to the right of the central figure show the distribution of manual inspection and estimated velocities, respectively.}
\label{fig:ansCorr}
\end{figure}

A similar analysis was performed on a second dataset derived from manual inspection of FF data products. For each spectrum analysed, a few $^{12}$CO features were identified within the list of features generated by the FF; the [NII] feature, if present, was used in the absence of easily identifiable $^{12}$CO emission. An average radial velocity was then calculated based on these selected features. As such, the velocity estimates within this dataset are obtained in a way that is substantially similar to the velocity estimating routine describe in this section. It should be noted, however, that not all $^{12}$CO features were identified in each spectrum, and this analysis was conducted using results from an earlier iteration of the FF script. This results in a slight methodological difference between the radial velocity estimating routine implemented in the FF and the manual inspection dataset, along with potential differences in the frequency position of the detected $^{12}$CO features. Consequently, additional uncertainties are introduced into the comparison, but the effect is expected to be small since modifications between FF script iterations did not significantly alter the detection of strong spectral features. A velocity estimate correlation plot for the manual inspection and FF datasets is shown in Fig.\,\ref{fig:ansCorr}. The average absolute value of the residual velocities in terms of the N flagging parameter is shown in Fig.\,\ref{fig:flagResidual} using green markers. The completeness curve for the absolute value of the residual velocities is shown in Fig.\,\ref{fig:Completeness}, again using green, with an inset showing the residual velocity distribution. The completeness curve and corresponding histogram use only estimates with \mbox{N $ > 3 $}. 

Estimates with a higher flagging parameter generally show better agreement with the comparison datasets. A slight exception to this trend is observed in the HIFI comparison for N = 7 and 9. Investigation shows that the discrepancy is due to individual outliers which inflate the error-bars for these points. The correlation plots and completeness curves for both real dataset comparisons show similar results. With the exception of estimates based on [NII], no high quality estimate deviated significantly from the comparison datasets. This deviation, however, is not due to a misidentification of the [NII] emission, but a consequence of estimating radial velocity using emission from different ISM components as discussed below. $\sim$90\% of estimates with \mbox{N $ > 3 $} have a residual velocity of 20\,\kms\ or less.

We combine both datasets, using the results based on manual inspection when the same source is present in both datasets, and consider only high quality estimates. This produces a sample size of 251 observations based on \CO rotational emission with a mean residual velocity of 5.8\,\kms\ and a standard deviation of 7.1\,\kms. The SPIRE FTS calibration paper \citep{Swinyard2014} measured a systematic frequency offset mapped into velocity space of $\sim$ 5\,\kms\ for expected line positions with a spread of $\sim$ 7\,\kms. As determined by our tests, the FF radial velocity estimation derived from $^{12}$CO detection has systematic and random errors consistent with these figures. For the subset of 49 observations with velocity estimates based on [NII], the mean residual velocity was 31.2\,\kms\ with a standard deviation of 102.1\,\kms. This higher discrepancy is due to 4 observations having residual velocities greater than 150\,\kms. These observations were inspected, and it was confirmed that the routine was correctly identifying the [NII] features. The discrepancy in velocities is thus attributed to the fact that [NII] is a tracer for more energetic regions of the ISM, whereas the dataset used for comparison contains velocity estimates mainly derived from lower energy emission. If these outliers are removed, the mean residual velocity becomes 4.1\,\kms\ with a standard deviation of 5.7\,\kms. 

\section{Cross-Correlation Routine}
\label{sec:xcor_routine}

The FF employs a secondary radial velocity estimating routine adapted from \cite{Zucker03}. The cross-correlation based routine estimates radial velocities of SPIRE observations by comparing synthetic spectra generated from the list of features detected by the FF, $\cal{F}$, and a line template, $\cal{T}$. The template contains 306 rest frame molecular and atomic lines commonly observed in FIR astronomical sources. This comparison is accomplished by synthesising a model spectrum for both the FF features and the line template, with the FF model spectrum shifted to a rest frame dependent on some assumed redshift. The two model spectra are then correlated, and this process is repeated in an iterative fashion over a range of assumed redshifts. The redshift which produces the highest correlation typically represents the redshift of the source under observation, though additional corrections are used in our routine. This process is described in more detail below.

To provide a valid comparison, the frequency sampling of the modelled spectra, $\mathcal{F}$ and $\mathcal{T}$, must be projected onto a common frequency axis. A frequency axis ranging from 400 to 1,600\,GHz using a step size of 0.1\,GHz was chosen. The model FF spectrum was generated by inserting detected features at their detected frequencies. After shifting these lines to a rest frame dependent on some assumed redshift, they were convolved with a Gaussian. To generate the model template spectrum, the template lines were inserted into an equivalent frequency axis and each feature was convolved with a Gaussian. The convolved Gaussian in both cases had an amplitude of 1, and a sigma of 0.05\,GHz (comparable to the frequency errors of detected features).

We scale $\mathcal{F}$ and $\mathcal{T}$, to meet the condition $\sum_{i} \mathcal{F}(f_i) = \sum_{i} \mathcal{T}(f_i) = 0$ used by the cross-correlation method in \cite{Zucker03}, and we calculate the standard deviation $ \sigma_\mathcal{F} $ and $ \sigma_\mathcal{T} $, for both model spectra. 

\begin{figure}
\centering
\vspace{-6pt}
\includegraphics[width=\columnwidth]{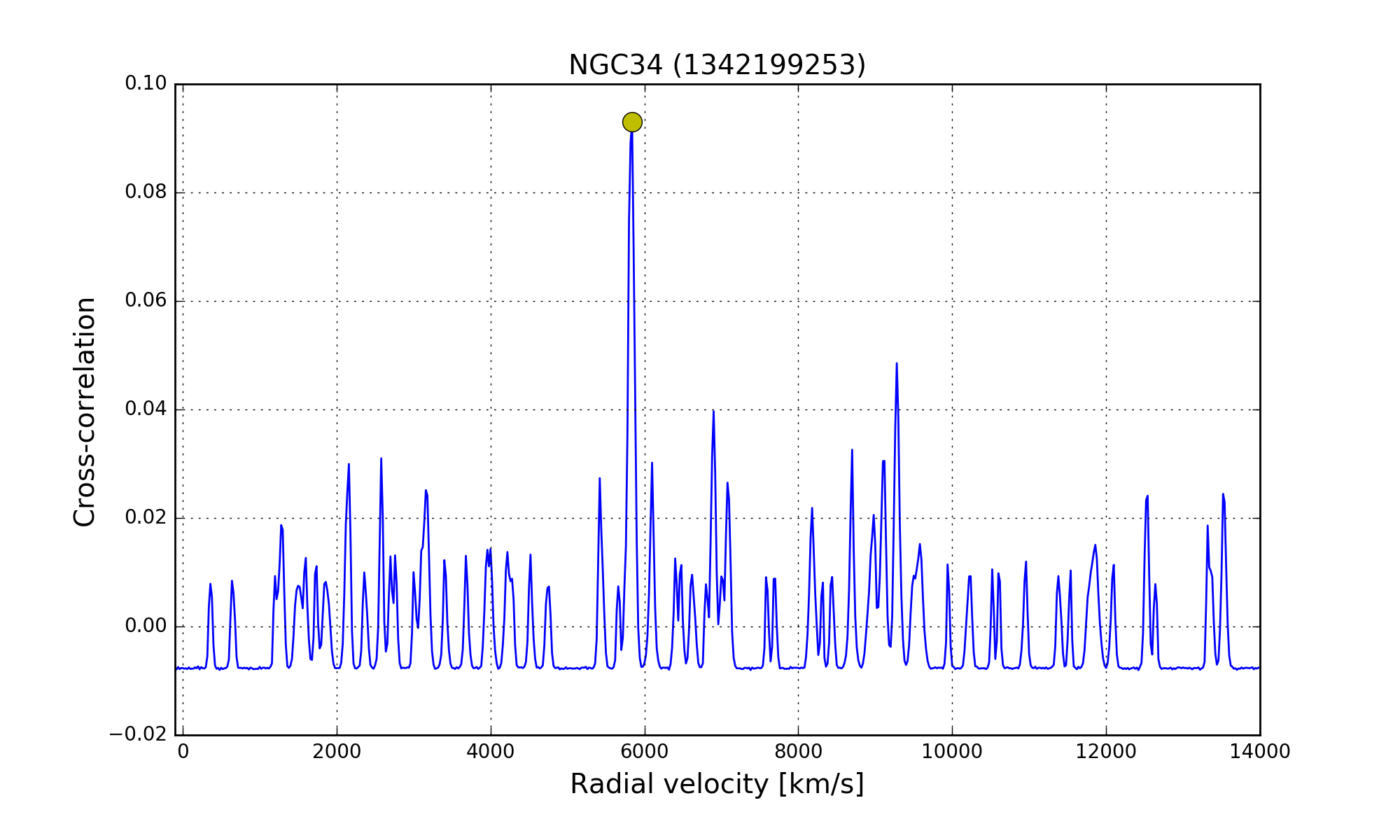}
\vspace{-18pt}
\caption{Cross-correlation result for NGC34 (\textsc{obsid}: 1342199253). Observation included 5 lines for SSW, and 5 for SLW. The yellow dot at 5840 \kms\ indicates the best radial velocity estimate.}
\label{fig_cs1}
\end{figure}

The initial model spectrum, $ \mathcal{F} $, is shifted to a rest frame model, $ \mathcal{F}_{0}(\mathrm{v}) $, by transforming the frequency axis using
\begin{equation}
f_{0} = \left(1 + \frac{\mathrm{v}}{c}\right) f \;,
\end{equation}
where $ c $ is the speed of light and $ \mathrm{v} $ is the assumed source radial velocity. The cross-correlation function is then 
\begin{equation}
\label{eq:cross-corr}
C(\mathrm{v}) = \frac{1}{N \sigma_\mathcal{F} \sigma_\mathcal{T}} \sum\limits_{f} \mathcal{T} \mathcal{F}_{0}(\mathrm{v}) \;.
\end{equation}
Eq.\,\ref{eq:cross-corr} is evaluated over the velocity range -1,000 to 14,000\,\kms\ in steps of 20\,\kms. An example of the cross-correlation function applied to NGC34 is shown in Fig.\,\ref{fig_cs1}. It can be seen that the correlation function reaches a maximum at a radial velocity of $ \sim 5,840 $\,\kms, corresponding to the radial velocity estimate of the source. 

The squared error of the radial velocity estimate was calculated using \cite{Zucker03}:
\begin{equation}
\label{eq_xcor_err}
\sigma_\mathrm{v}^2 =- \Big[N \frac{C''(\hat{\mathrm{v}})}{C(\hat{\mathrm{v}})} \frac{C^{2}(\hat{\mathrm{v}})}{1-C^{2}(\hat{\mathrm{v}})}\Big]^{-1} \;,
\end{equation}
where ${C''(\mathrm{v})}$ is the second derivative of the cross-correlation function and $ \mathrm{\hat{v}} $ is the velocity estimate which maximises the cross-correlation function.

While performing the cross-correlation, we skipped features flagged as a poor fit in a noisy region \citepalias{FFtech}. Additionally, we chose only lines with SNR above 10, provided at least 5 of them exist in the observation. Otherwise, we chose only features with SNR above 9, and so on down to a minimum SNR of 5. If all of the lines in the observation were faint, we performed the cross-correlation on the full set of lines except those flagged as a poor fit in a noisy region. This process is used to maintain a high quality cross-correlation by reducing the possibility of spurious features being used in the analysis.

\begin{figure}
\centering
\vspace{-6pt}
\includegraphics[width=\columnwidth]{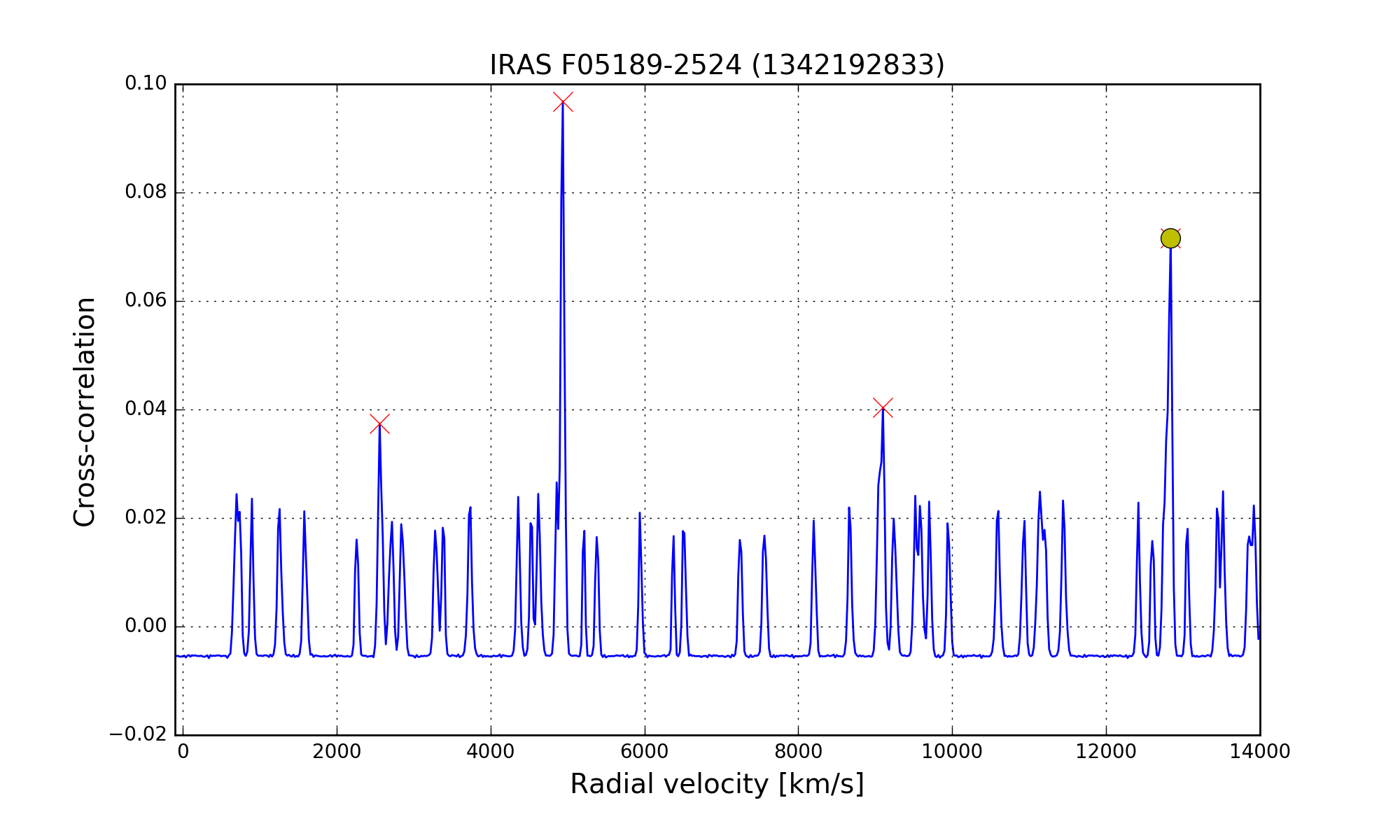}
\vspace{-15pt}
\caption{Cross-correlation with multiple possible peaks for IRAS F05189-2524 (\textsc{obsid}: 1342192833). The red crosses show the significant peaks, while the yellow dot at 12840 km s$^{-1}$ indicates the peak which maximizes the number of identified lines.}
\label{fig_cs2}
\end{figure}

Many far-infrared sources express the \CO ladder which contains a number of lines that are separated by a fixed frequency. This creates degeneracy in the cross-correlation, and as a result, we may have more than one significant peak for $C(\mathrm{v})$. As such, the maximum correlation value in some observations may not necessarily indicate the correct velocity estimate. To reduce this problem, we extended the cross-correlation method with line identification applied to the five most significant peaks within the cross-correlation function. While performing line identification, we compared the rest frequencies of all observed lines, even those previously omitted due to their SNR or feature flag, with the input template. The line is considered to be identified when the difference between a line in $ \mathcal{T} $ and $\mathcal{F}_{0}$ is less than 0.3\,GHz. If multiple features are within 0.3\,GHz of the template line, the feature which minimises this difference is used. The cross-correlation peak that maximises the number of identified lines is considered the best velocity estimate, $\bar{\mathrm{v}}$ (see Fig.\,\ref{fig_cs2}). 

Problems may also occur due to the finite frequency sampling and the discrete velocity sampling resulting in situations where a few $\cal{F}$ lines overlapping with the template $\cal{T}$ can give the misleading impression of a very high cross-correlation value while many lines remain unidentified. This may happen for observations containing very few features, many spurious detections, or significant frequency errors in the FF detected lines.

\subsection{Improving the radial velocity}
\label{impr_xcor} 
Once we have the radial velocity estimate, $\bar{\mathrm{v}}$, for a given observation derived from the steps outlined above, we shift the detected FF lines to the rest-frame and performed line identification with the template as previous described. With the purpose of improving the velocity estimate by removing any systematic bias, we calculate the velocity equivalent offset between the identified FF rest-frame lines $ \mathcal{F}_0 $ and the template lines $ \mathcal{T} $,  $\Delta \mathrm{v_{ID}}$,  and subtract its median from the radial velocity estimate to obtain:
\begin{equation}
\label{eq:xcor_improved}
\mathrm{v}_\mathrm{xcor} = \bar{\mathrm{v}} - \mathrm{median}(\Delta \mathrm{v_{ID}}) \;.
\end{equation}

We calculate the error on the improved radial velocity $\mathrm{v}_\mathrm{xcor}$ as the sum in quadrature of the cross-correlation velocity error (see Eq.\,\ref{eq_xcor_err}) and the robust error from $\Delta \mathrm{v_{ID}}$:
\begin{equation}
\sigma^2_{\mathrm{vxcor}} = \sigma_\mathrm{v}^2 + \left[1.4826 \times \mathrm{MAD}(\Delta  \mathrm{v_{ID}})\right]^2 \;,
\label{eq_rvel_err}
\end{equation}
where and MAD represents the Median Absolute Deviation, the factor 1.4826 converts it to an estimator of the standard deviation. In almost all cases the error from $\Delta \mathrm{v_{ID}}$ dominates.

In the cross-correlation phase some of the faint features or features with a poor fit in a noisy region of the spectrum, were omitted. We reiterate that line identification uses all features detected by the FF, even if they were previously flagged out in the cross-correlation calculation. 

\subsection{Radial velocity for sources with too few lines}

The cross-correlation routine is unreliable if $\mathcal{F}$ contains too few features. In practice, for $\mathcal{F}$ observation catalogue entries with less than four features, the radial velocity was estimated by searching for the line with the highest SNR and assumed that it is either [NII] at 205.1 $\mu$m (1461.13\,GHz) if it appears in the SSW band, or \CO(J=7-6) at 806.65\,GHz if it appears in the SLW band. These assumptions are generally valid as the [NII] line is usually the strongest line in the spectra of many galactic and extragalactic sources. Similarly, the \CO(J=7-6) line is often one of the brightest \CO lines in molecular clouds. In some cases these assumptions are wrong and we manually inspect the results for all sources with too few lines.

\subsection{Validation}
\label{subsec:Validation_xcor}

\begin{figure}
\centering
\vspace{-6pt}
\includegraphics{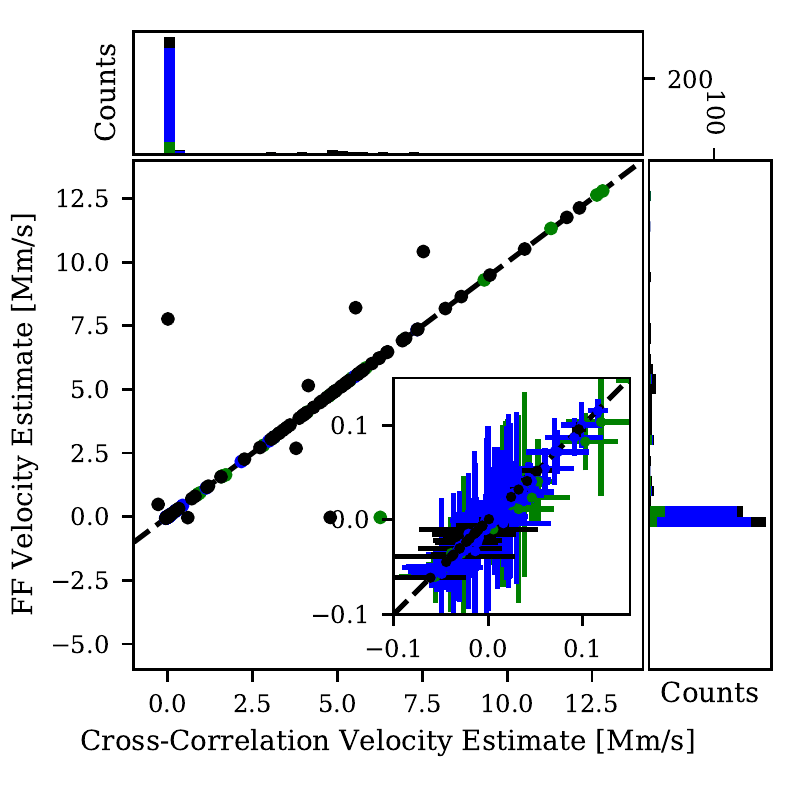}
\vspace{-15pt}
\caption{Correlation between radial velocity estimates derived through the \CO based routine meeting the FF selection criteria, and the cross-correlation based routine for complementary observations. The colour scheme used is the same as in Fig.\,\ref{fig:simCorr}. Estimates derived from [NII] emission are indicated with black markers. The inset shows a close-up of the lower velocity cluster. Histograms on top and to the right of the central figure show the distribution of cross-correlation and $^{12}$CO derived velocity estimates, respectively. Only estimates based on [NII] features deviate significantly from the one-to-one line.}
\label{fig:Xcorr}
\end{figure}

To assess the reliability of the cross-correlation based routine, we compare the high quality radial velocity estimates derived through the \CO based routine with the velocity estimates from the cross-correlation routine for the same sources. The resulting comparison is shown in Fig.\,\ref{fig:Xcorr}. Of the 423 estimates compared, only 12 showed an absolute residual velocity greater than 20\,\kms, representing a 97\% success rate. The estimates with high residuals were universally derived from potential [NII] emission using the \CO based routine. The corresponding spectra express few detected features which makes estimating radial velocity intrinsically difficult. Manual inspection of the 12 sources with high residuals reveals that what appears to be [NII] features were correctly identified by the \CO based routine. As such, estimates derived through the \CO based routine take precedence over those derived through the cross-correlation based routine. However, given the agreement between the two routines demonstrated in Figure\,\ref{fig:Xcorr}, estimates dervided through the cross-correlation based routine are expected to be fairly reliable.

\section{External References}
\label{sec:externalRef}

We consulted external references in an attempt to obtain velocity estimates for observations which did not yield reliable radial velocity estimates from either the \CO or cross-correlation based routines. These references include:
\begin{itemize}
\itemsep-1em 
\item The aforementioned complimentary HIFI velocity estimate dataset \citep{hifiVelocity}. \\
\item The Set of Identifications, Measurements, and Bibliography for Astronomical Data (SIMBAD; \cite{wenger2000simbad}). \\
\item An analysis of dusty star-forming galaxies using SPIRE by \cite{wilson17}. 
\end{itemize}
Velocity estimates derived from the above sources are denoted with flags in the FF catalogue by `H?', `S?', and `W17', respectively. External references were used only for the SPIRE sparse observation catalogue, and only when the references contained estimates for a source residing within 6" of the SPIRE observation's nominal coordinates.

\section{Feature Finder Results}
\label{sec:FF_Results}

An attempt was made to estimate radial velocity for all HR SPIRE FTS observations processed by the FF. Results as they pertain to mapping observations and the central detectors of sparse observations are presented here. The extension to off-axis detectors of sparse observations is presented in a companion paper \citep[\citetalias{FFlineID}:][]{FFlineID}. 

\begin{figure*}
\centering
\vspace{-15pt}
\includegraphics[width=1.0\linewidth]{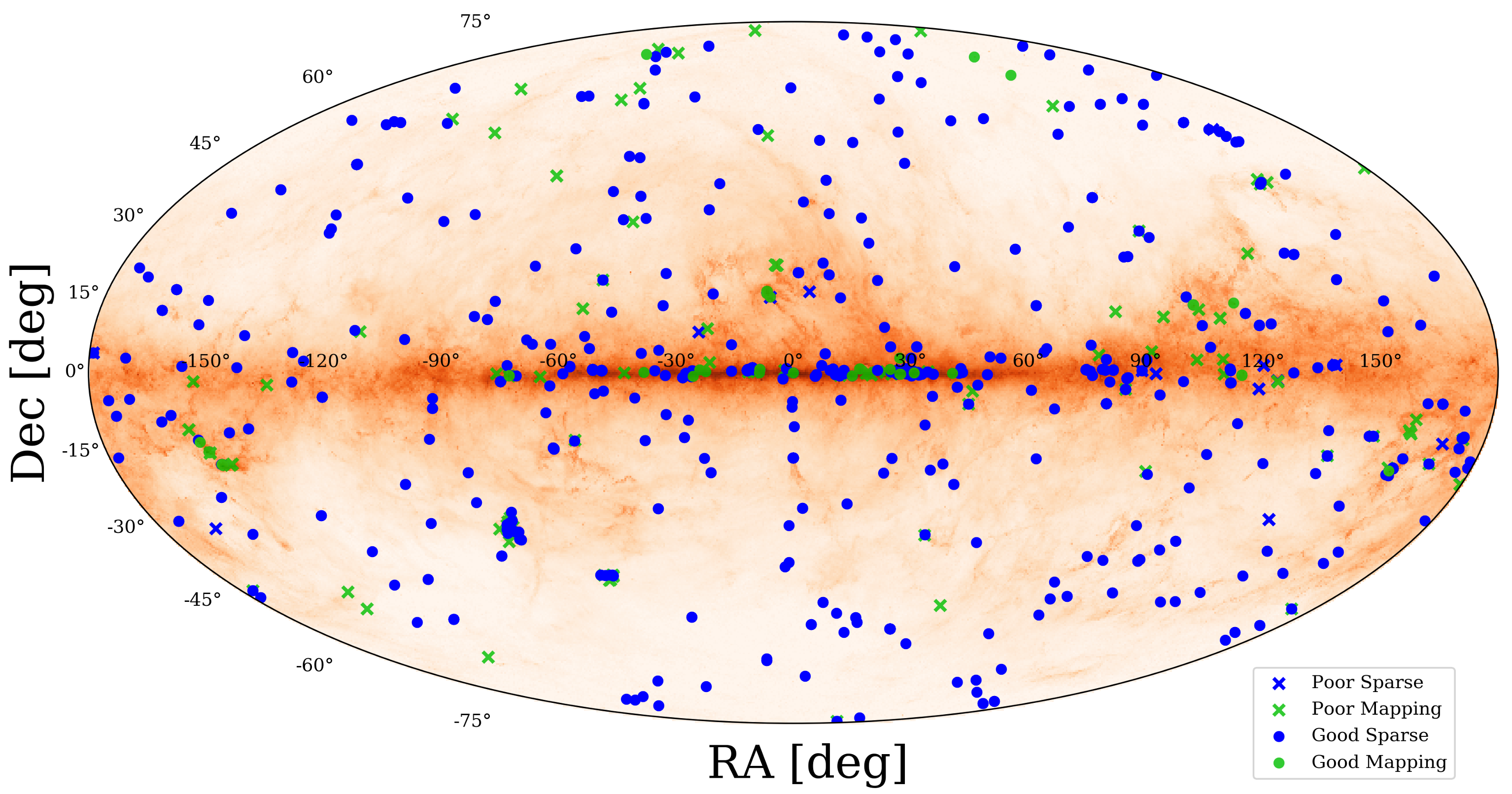}
\vspace{-15pt}\caption{On-sky projection of observations processed by our radial velocity estimating routine. Green and blue markers indicated SPIRE FTS mapping and sparse observations, respectively. Mapping observations with at least half of the feature containing pixels receiving a velocity estimate, and sparse observations receiving a velocity estimate are indicated with circles. Observations failing this criteria are indicated with crosses. The image background is a composite map using Planck \CO and integrated 857\,GHz data obtained from NASA's IRAS archive \protect\footnotemark.}
\label{fig:FullSky}
\end{figure*}

An on-sky projection of these sources is shown in Fig.\,\ref{fig:FullSky}. Markers are colour-coded based on observation mode with mapping and sparse observations receiving green and blue markers, respectively. `Good' velocity estimates for sparse observations, those which meet the FF flagging criteria, are indicated with circles; crosses are used otherwise. Mapping observations contain many pixels, each receiving a unique velocity estimate. Furthermore, not every pixel in the observation will contain a spectrum with detectable features, and thus, would not be expected to produce a valid velocity estimate. As such, we adopt a heuristic criteria where mapping observations are considered `good' if half or more pixels with detected features receive a velocity estimate meeting the FF flagging criteria. Good mapping observations are indicated with circles; crosses are used otherwise. 

\begin{table*}
\begingroup
\begin{center}
\newdimen\tblskip \tblskip=5pt
\caption{\label{tab:redshift}Distribution of radial velocity estimating methods by observation mode/calibration. For sparse observations, a velocity is estimated for the spectrum derived from the overlapping central detectors of the SLW and SSW arrays. For mapping observations, a unique velocity estimate is obtained for individual pixels in the spectral map.}
\nointerlineskip
\small
%
\newdimen\digitwidth
\setbox0=\hbox{\rm 0}
\digitwidth=\wd0
\catcode`*=\active
\def*{\kern\digitwidth}
\newdimen\signwidth
\setbox0=\hbox{+}
\signwidth=\wd 0
\catcode`!=\active
\def!{\kern\signwidth}
%
\tabskip=2em plus 2em minus 2em
\halign to \hsize{
\hfil**********#&
**#\hfil& 
\hfil#\hfil& 
\hfil#\hfil& 
\hfil#\hfil& 
\hfil#\hfil& 
\hfil#\hfil& 
\hfil#\hfil& 
\hfil#\hfil& 
\hfil#**\hfil&
#**********\hfil\cr
 &\multispan9\hrulefill& \cr
\noalign{\vspace{-8.0pt}}
 &\multispan9\hrulefill& \cr
& & FF& FF?& XCOR& XCOR?& H?& S?& W17& Total& \cr
\noalign{\vspace{-5.5pt}}
 &\multispan9\hrulefill& \cr

 & Sparse pnt& 205& **\,226& 15& 147& 1& 90& 58& **\,742& \cr 
 & Sparse ext& *53& **\,184& *9& *86& 0& 17& *6& **\,355& \cr 
 & Mapping&    **0& 22\,937& *0& **0& 0& *0& *0& 22\,937& \cr 
\noalign{\vspace{-5.5pt}}
 &\multispan9\hrulefill& \cr
}
\end{center}
\endgroup
\vspace{-12pt}
\end{table*}

For SPIRE point source calibrated sparse observations, a total of 742 radial velocity estimates were obtained; this represents $ \sim $91\% of all observations processed. For SPIRE sparse observations processed using extended source calibration, 355 radial velocity estimates were obtained which again represents $ \sim $91\% of all observations processed. For mapping observations, only $ \sim $31\% of observations met the `good' criteria outlined above and indicated in Fig.\,\ref{fig:FullSky}. Using an alternative metric, 22,937 mapping pixels received a velocity estimate which represents $ \sim $42\% of all mapping pixels containing detected features. These results are summarized in Tab.\,\ref{tab:redshift} which also indicates the distribution of velocity estimating methods used. Estimates based on the \CO method are denoted by `FF?' while estimates derived using the cross-correlation method are denoted by `XCOR?'. These estimates were compared with the companion HIFI velocity estimate dataset. If a suitable comparison velocity estimate is found, and if the difference between the two velocities is $ < $20\,\kms\ or $ < $20\% if the velocity estimate is $ > $100\,\kms, the velocity flags are modified to `FF' or `XCOR', respectively. In the FF sparse catalogue, the method used in deriving the radial velocity estimate is presented in the metadata of each catalogue under the keyword `FLAG\_RV'.

\begin{figure}
\centering
\begin{minipage}{1.0\columnwidth}
\includegraphics[width=1.0\linewidth]{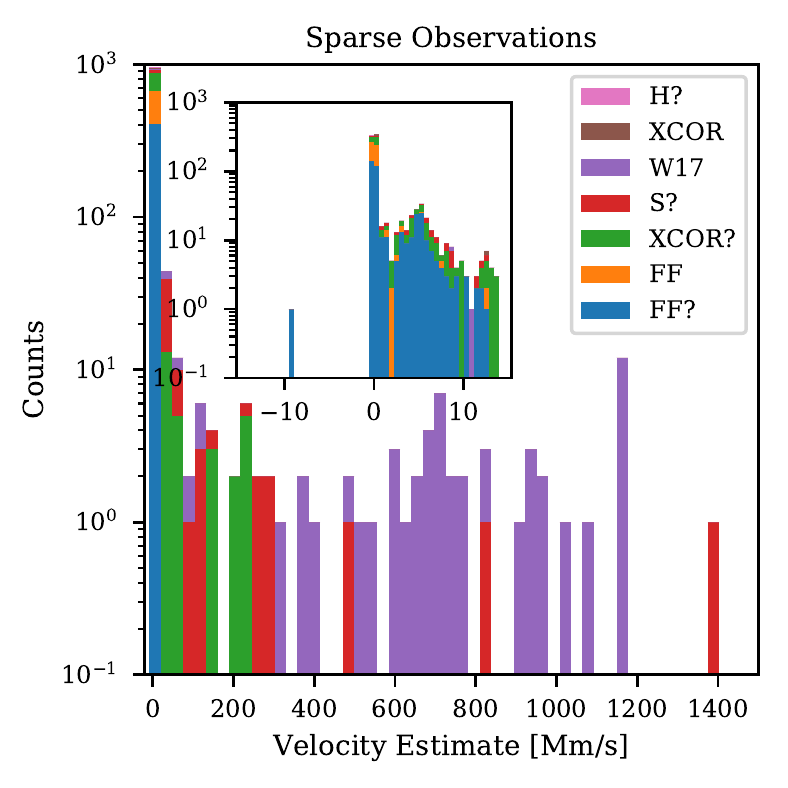}
\end{minipage}
\begin{minipage}{1.0\columnwidth}
\includegraphics[width=1.0\linewidth]{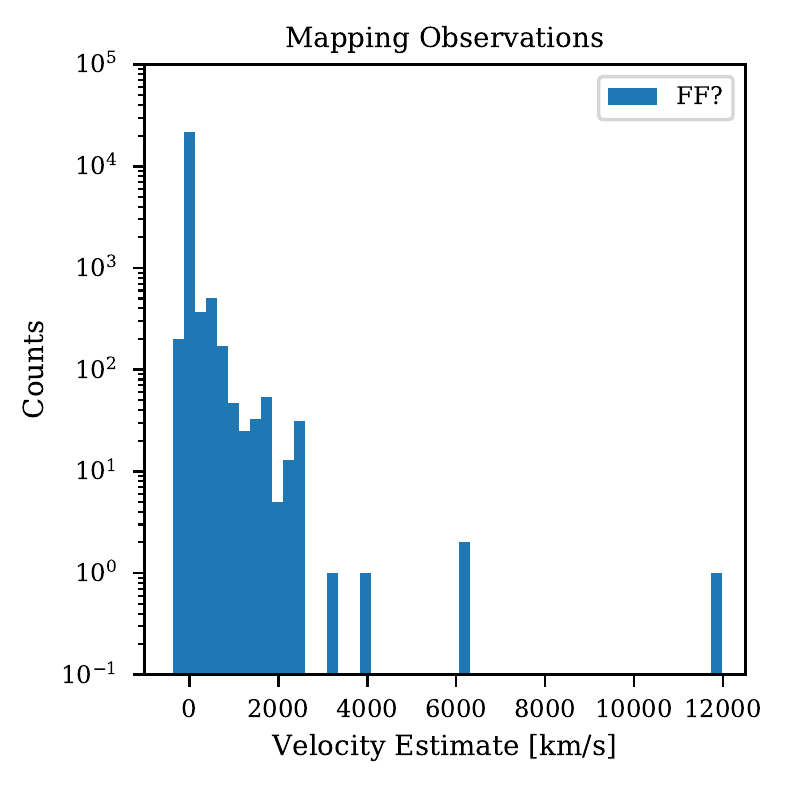}
\end{minipage}
\caption{Radial velocity estimates for sparse (top) and mapping (bottom) observations in terms of velocity estimating method used. The inset in the sparse observation frame shows a close-up of the -10,000 to 15,000\,\kms\ range of the main plot. The sparse observations plot combined the velocity estimates from both the point-source and extended-source calibrated catalogues.}
\label{fig:ObsVelocityHistogram}
\end{figure}

The distribution of radial velocity estimates is shown in Fig.\,\ref{fig:ObsVelocityHistogram}. The sparse observation dataset contains 306 extra-galactic sources, as indicated by the HSA, which would reasonably have high radial velocities, while the mapping dataset contains only galactic sources with typically lower radial velocities. This helps to explain the difference observed in the velocity distributions between the two datasets.

In some cases, SPIRE observed the same source multiple times. This was the case for the spectral calibration sources \citep{Hopwood15}, for example. As a consistency check, radial velocity estimates from redundant observations, those which reside within a 6" radius on-sky, were inspected. It was confirmed that all such groups expressed similar velocity estimates as would be expected. 

\section{Discussion}
\label{sec:discussion}

When using the \CO based velocity estimating routine, accurate estimates are in general dependent on the presence of strong $^{12}$CO features. Spectra with weak $^{12}$CO emission and an abundance of other prominent features tend to result in poor estimates, and this inaccuracy is expected to increase with source velocity. However, the routine seems to be resilient in term of the number of lines used as long as significant $^{12}$CO features are present. Using real FF products, the velocity estimate for all 90 spectra with 40 or more detected features were examined. Only 5 sources had poor velocity estimates, one of which was correct, but did not meet the baseline number of candidate features acceptance threshold. This results in a 94.4\% success rate for sources with a large number of features. The 4 sources with inaccurate velocities have spectra where the $^{12}$CO features, if present, are difficult to identify even by manual inspection, and exhibit prominent features which are not $^{12}$CO. 

A critical limit of this routine is the upper limit on radial velocity. Since the routine pairs identified $^{12}$CO candidate features to their nearest $^{12}$CO rest frequency emission, significantly shifted features will not be paired correctly. Based on our minimum flagging criteria of identifying 7 $^{12}$CO candidate features, this breakdown occurs at around $ \pm $14,000\,\kms.

The N flagging parameter, representing the number of $^{12}$CO candidate features used when estimating radial velocity, appears to be a robust indicator of the accuracy of velocity estimates. In our simulation, although a minority of cases occur where accurate velocity estimates are rejected ($\sim$16\%), no cases of poor velocity estimates with good flagging values have been observed either in real or simulated data. Estimates based on the ionized nitrogen line show the greatest deviation from comparison datasets, but this is a result of [NII] probing more energetic components of the ISM than rotational \CO emission, and does not appear to be the result of incorrectly identified [NII] lines.

The cross-correlation based routine is a good compliment to the \CO based routine in that it is capable of estimating radial velocity in the absence of strong \CO emission. One of the main complications in employing the cross-correlation based routine results from the degeneracies in the correlation function due to the \CO ladder, and conveniently enough, it is these degeneracies which the \CO based routine uses to derive its velocity estimates. Common to both routines is the limitations associated with estimating the redshift of spectra which express few spectral features. In these cases, both routines attempt to identify potential [NII] emission. Our tests show that both routines are reasonably capable of identifying [NII] when it is present, though in cases where the routines disagree, it appears that identification using the \CO based method is more accurate.

When both routines fail to produce a valid velocity estimate, we consult external references. Though it is expected that these external references are reliable, it should be emphasised that radial velocities derived from different emission features may produce different velocity estimates as these emission features may be produced by different components of the same source. That is the warmer regions of a source emitting at optical frequencies may have different dynamics than the cooler molecular clouds emitting at FIR frequencies. 

\section{Conclusions}
\label{sec:Conclusion}

We constructed two radial velocity estimating routines. The first is based on the identification of shifted rotational $^{12}$CO emission in SPIRE spectra. Identification is accomplished by searching frequency differences between features detected by the FF for the characteristic differences of the $^{12}$CO ladder to within a velocity dependent tolerance. Features with the highest number of matches are most likely to be $^{12}$CO and are chosen as potential \CO candidate features. These candidate features are paired with their nearest \CO rest frequency emission line. Multiple $^{12}$CO candidate features paired to the same rest frequency are filtered by their SNR with all but the highest SNR feature being rejected. Features which produce high standard deviations in the calculated velocities are removed, making the routine somewhat resistant to false \CO emission identification. The total number of resulting $^{12}$CO candidates, `N', is used to assess the quality of the estimate. If no $^{12}$CO is found, we attempt to estimate source velocity by searching for a high SNR feature within 60\,GHz of the [NII] rest frequency. Estimates based on candidate features with fewer than two matches are rejected while estimates  with N > 6, or based on [NII], are included in the FF catalogue SAFECAT. This criteria is lowered to N $ > $ 3 for mapping observation pixels contain only SLW spectra.

The second routine is based on the cross-correlation of features identified by the FF and a line template containing most of the molecular and atomic lines commonly observed within the FIR. Model spectra are constructed for both the FF products and the line template. The FF model spectrum is corrected for some assumed redshift and the two model spectra are correlated. The redshift which produces the highest correlation is typically assumed to be the redshift of the source. The corresponding velocity estimate is checked using feature identification of the five most prominent peaks in the cross-correlation function. The peak which maximises the number of identified lines is takes and the best estimate of the radial velocity. The estimate is further refined by subtracting the median velocity difference between the identified features and the corresponding template features. When spectra express few features, velocity estimates are derived using the \CO(J=7-6) and [NII] lines by searching for the most prominent feature in the SLW and SSW bands, respectively. 

Tests using real and simulated data demonstrate the reliability of our routines. It has been observed that $ \sim $90\% of velocity estimates are within $ 20 $\,\kms\ of the velocity estimates derived from comparison datasets. The most significant deviations occur with estimates based on [NII] emission, however, this deviation is attributed to [NII] probing different components of the ISM than \CO emission and does not appear to be the result of falsely identified [NII] features. 

By employing both routines, the most significant limitation is when spectra express few spectral features making redshift determination intrinsically difficult. When both routines fail to produce valid velocity estimates, external references are consulted (see \S\,\ref{sec:externalRef}). 

In total, velocity estimates were derived for $ \sim $91\% of HR SPIRE sparse observations and $ \sim $42\% of mapping pixels with detected features.

\footnotetext{https://irsa.ipac.caltech.edu/data/Planck/release\_2/all-sky-maps}

\section*{Acknowledgements}

\emph{Herschel} is an ESA space observatory with science instruments provided by European-led Principal Investigator consortia and with important participation from NASA\@. This research acknowledges support from ESA, CSA, CRC, CMC, and NSERC\@. 

This research has made use of the NASA/IPAC Infrared Science Archive, which is funded by the National Aeronautics and Space Administration and operated by the California Institute of Technology.

This research has made use of the SciPy (\url{www.scipy.org}) and Astropy (\url{www.astropy.org}) Python packages. Table formatting in this paper followed the {\it Planck} Style Guide \citep{PlanckStyle}. 
\nocite{2020SciPy-NMeth, astropy:2013, astropy:2018}\\

\section*{Data Availability}
The \textit{Herschel} SPIRE Spectral Feature Catalogue has been assigned an ESA Digital Object Identifier (DOI) and is available at: \href{https://doi.org/10.5270/esa-lysf2yi}{doi.org/10.5270/esa-lysf2yi}. The FF code and all FF products are publicly available via the \textit{Herschel} Science Archive. 




\bibliographystyle{mnras}
\bibliography{mnras_Redshift_revised}







\bsp	
\label{lastpage}
\end{document}